\documentclass[a4paper,11pt]{article}%
\usepackage{amsfonts}
\usepackage{graphicx}
\usepackage{amsmath}
\usepackage{amssymb}%
\usepackage{cite}
\pdfoutput=1
\makeatletter

\@addtoreset{equation}{section}
\makeatother
\newcommand{\beq}{\begin{equation}}
\newcommand{\eeq}{\end{equation}}
\renewcommand{\a}{\alpha}
\renewcommand{\b}{{{\beta}}}

\newcommand{\be}{\begin{eqnarray}}
\newcommand{\ee}{\end{eqnarray}}

\begin{document}
\baselineskip=18pt
\baselineskip 0.6cm
\begin{titlepage}
\setcounter{page}{0}
\renewcommand{\thefootnote}{\fnsymbol{footnote}}
\begin{flushright}
YITP-17-114\\
ARC-17-10
\end{flushright}
\begin{center}
{\Large \bf $\mathrm{D=10}$ Super-Yang-Mills Theory and \\ Poincar\'e
Duality in Supermanifolds }\\
\vskip 0.5cm
{\large Pietro
Fr\'e$^{~a,c,d,e}$\footnote{pfre@unito.it} and Pietro Antonio
Grassi$^{~b,c,d}$\footnote{pietro.grassi@uniupo.it} } \vskip 0.2cm {
\small \centerline{$^{(a)}$ \it Dipartimento di Fisica, Universit\`a
di Torino,} \centerline{\it via P. Giuria , 1, 10125 Torino, Italy.}
\medskip
\centerline{$^{(b)}$ \it Dipartimento di Scienze e Innovazione
Tecnologica,} \centerline{\it Universit\`a del Piemonte Orientale,}
\centerline{\it viale T. Michel, 11, 15121 Alessandria, Italy.}
\medskip
\centerline{$^{(c)}$ \it INFN, Sezione di Torino,} \centerline{\it
via P. Giuria 1, 10125 Torino.}
\medskip
\centerline{$^{(d)}$ \it Arnold-Regge Center,}
\centerline{\it via P. Giuria 1,  10125 Torino, Italy}
\medskip
\centerline{$^{(e)}$\it  National Research Nuclear University MEPhI
(Moscow Engineering Physics Institute),} \centerline{\it Kashirskoye
shosse 31, 115409 Moscow, Russia } }
\end{center}
\vskip 0.2cm
\centerline{{\bf Abstract}}
\medskip
\noindent { We consider super Yang-Mills theory on supermanifolds
$\mathcal{M}^{(D|m)}$ using integral forms. The latter are used to
define a geometric theory of integration and are essential for a
consistent action principle. The construction relies on Picture
Changing Operators $\mathbb{Y}^{(0|m)}$, analogous to those
introduced in String Theory, that admit the geometric interpretation
of Poincar\'e duals of closed submanifolds of superspace
$\mathcal{S}^{(D|0)} \subset \mathcal{M}^{(D|m)}$ having maximal
bosonic dimension $D$. We discuss the case of Super-Yang-Mills
theory in $D=10$ with $\mathcal{N}=1$ supersymmetry and we show how
to retrieve its pure-spinor formulation from the rheonomic
lagrangian $\mathcal{L}_{rheo}$ of D'Auria, Fr\'e and Da Silva,
choosing a suitable $\mathbb{Y}^{(0|m)}_{ps}$. From the same
lagrangian $\mathcal{L}_{rheo}$, with another choice
$\mathbb{Y}^{(0|m)}_{comp}$  of the PCO, one retrieves the component
form of the SYM action. Equivalence of the formulations is ensured
when the corresponding PCO.s are cohomologous, which is true, in
this case, of $\mathbb{Y}^{(0|m)}_{ps}$ and
$\mathbb{Y}^{(0|m)}_{comp}$.}
\end{titlepage}
\setcounter{footnote}{0} \newpage\setcounter{footnote}{0}
\section{Introduction}
In 10 dimensions, there are few examples of supersymmetric models
which are consistent and well-defined. Among them, there are
super-Yang-Mills theory with $\mathcal{N}=1$ supersymmetry (sixteen
supercharges), and three supergravity theories: $\mathcal{N}=1$ type
I, $\mathcal{N}=2$ type IIA and $\mathcal{N}=2$ type IIB. This is
essentially due to the fact that, in 10 dimensions, the spinorial
representation of the Lorentz group $\mathrm{SO(1,9)}$ admits a
minimal realization in terms of Majorana-Weyl spinors with two
different chiralities (A and B). Therefore only the massless
physical multiplets with a spin-one vector field (super-Yang-Mills)
or the spin-two graviton constitute closed multiplets with respect
to the action of the supercharges \cite{Nahm:1977tg}. From the
string theory side, we have learnt that the lowest massless states
of the open superstring theory and of the closed superstring theory
are respectively described by those super gauge theories and
supergravities \cite{Green:2012oqa,Green:2012pqa}.
\par
Here we consider only the case of super-Yang-Mills in 10 dimensions
with $\mathcal{N}=1$ supersymmetry. That theory is characterized by
a supermultiplet containing the gauge degrees of freedom (eight
on-shell d.o.f.'s) and those of its supersymetric partner, the
gaugino (eight on-shell d.o.f.'s). The off-shell theory is described
by a component action which is manifestly invariant under Lorentz
and gauge symmetry and supersymmetric up to total derivatives
\cite{Brink:1976bc}. Moreover, there is a superspace formulation of
the theory \cite{Witten:1985nt,Gates:1986tj} which is valid only for
on-shell fields. Namely, in order to take into account the correct
supersymmetry transformations and the quantum numbers, the gauge
field and the gaugino are described by a single superconnection
(which is a one-superform with a vectorial and a spinorial
superfield). Unfortunately, in order to match the physical degrees
of freedom one has to choose specific constraints on the
superconnection and, at the end of the day, these latter imply also
the equations of motion. Hence,   there seems to be no off-shell
superspace formulation that leads to a superspace action (some
attempts and remarks can be read in \cite{Berkovits:1997wj}) or, put
differently, there is only an on-shell superspace formulation
\cite{Harnad:1985bc}.
\par
The needed superspace constraints to be imposed on the
superconnections can be derived as integrability conditions for the
motion of a superparticle. This  has been used in
\cite{Witten:1985nt,Howe:1991mf,Howe:1991bx} and it has been
successfully implemented at the string level in
\cite{Berkovits:2000fe,Berkovits:2001rb}. It has been shown that
BRST cohomology restricted to the shell of pure spinor constraints
coincides with the superstring spectrum. In the case of open
superstrings, restricting the latter construction to the massless
sector one obtains a pure-spinor formulation of D=10 $\mathcal{N}=1$
SYM. The success of the pure spinor approach as a supersymmetric
target space formulation of superstrings prompted the analysis of
scattering amplitudes and quantum corrections to classical dynamics.
Indeed such effects can be mastered only by means of an effective
action. More significantly, an effective off-shell action is needed
for any string field theory analysis. It has been shown that, upon
integration over the bosonic coordinates, over the fermionic
coordinates and over the pure-spinor zero modes, a Chern-Simons type
Lagrangian  provides a convenient reformulation of super-Yang-Mills
theory within a superspace framework. Such a formulation appears to
be the so far known  presentation of the theory that  is the closest
to a true superspace action of D=10 $\mathcal{N}=1$ SYM.
\par
From a completely different point of view, many years ago the Torino
group established a formulation of supersymmetric field theories in
a more geometrical setup \cite{castdauriafre}. In such an approach
each individual field of the space-time component formulation is
promoted to the status of a superfield and  the large amount of
non-physical components is suppressed by a collection of suitable
constraints. Furthermore, a variational principle is provided that
yields, just in one stroke, both the physical equations of motion
and the suitable superspace constraints mentioned above.
Unfortunately this variational principle is formulated in a very
intrinsic way by specifying the immersion of the bosonic space-time
into superspace; in the case of interest to us here, it is the
immersion of a ten-dimensional bosonic manifold into the $(10|16)$
supermanifold that is at  stake. In such a general setup, that is
dubbed the {\it rheonomic formulation}, the variation of the
immersion is compensated by a diffeomorphism of the action. In our
case the action is a suitable 10-form that was derived many years
ago by D'Auria, Fr\'e and da Silva \cite{DAuria:1981zjr},
generalizing similar constructions of SYM theories in D=4 pioneered
by Fr\'e. \cite{Fre:1981ny,Fre:1981my}. One of the main advantages
of this formulation  is that it naturally provides the coupling of
the rigid supersymmetric theory to supergravity or its localization
on curved supermanifolds, such as coset superspaces.
\par
Recently in \cite{Castellani:2014goa,Castellani:2015paa}, an
alternative formulation, based on the rheonomic action, was
discovered. The immersion of the bosonic submanifold into the
supermanifold has been implemented introducing suitable
\textit{Poincar\'e duals} in superspace.   This  leads to an
\textit{integral form} suitable to be integrated over the supermanifold.
The choice of the Poincar\'e dual amounts to choose a specific
representation of the theory. In several papers
\cite{Castellani:2017ycm,pappo1,Castellani:2016ibp,Grassi:2016apf},
utilizing  different choices of the Poincar\'e dual (in the text the
form will be denoted by {\it PCO}), it has been shown how one can
interpolate between the several formulations of the same
supersymmetric model that range from the component action to the
superspace action.
\par
Here, we  show that by the same token  we are able to interpolate
between the component action of D=10 $\mathcal{N}=1$ super
Yang-Mills and the pure spinor action of the same theory 
\cite{Berkovits:2000fe,Berkovits:2001rb}. In all
cases, the differential form that is paired with the Poincar\'e dual
of a  suitable 10-dimensional cycle in superspace $\mathcal{C}
\subset \mathcal{M}^{10|16}$ is the rheonomic Lagrangian of
\cite{DAuria:1981zjr}. Schematically one has:
\begin{equation}
\begin{array}{lclcl}
  \mbox{Action of formulation A} &=& \int_{\mathcal{C}_A} \, \iota^\star\left[\mathcal{L}_{rheonomic}\right] &=&
  \int_{\mathcal{M}^{10|16}} \mathcal{L}_{rheonomic} \wedge \mathbb{Y}^{0|16}_A   \\
  \mbox{Action of formulation B} &=& \int_{\mathcal{C}_B} \, \iota^\star\left[\mathcal{L}_{rheonomic}\right] &=&
  \int_{\mathcal{M}^{10|16}} \mathcal{L}_{rheonomic} \wedge \mathbb{Y}^{0|16}_B
  \end{array}
  \label{chudo}
\end{equation}
where
\begin{equation}\label{cortedeimiracoli}
    \iota \quad : \quad \mathcal{C}_{A,B} \, \hookrightarrow \, \mathcal{M}^{10|16}
\end{equation}
is the immersion map and $\iota^\star\left[\dots\right]$ denotes its
pull-back. The picture changing operators $\mathbb{Y}^{0|16}_{A,B}$
are the Poincar\'e duals of the corresponding cycles.
\par
Since homology theory in superspace is conceptually and technically
very hard and ill-defined, the analogue of a simplicial approach
being missing, the dual cohomological formulation turns out to be
much more promising and it is well founded on the theory of
\textit{integral forms}. Eventually integral forms
$\mathbb{Y}^{0|16}$ define their Poincar\'e dual cycles and
implicitly define the immersion maps (\ref{cortedeimiracoli}). The
various pull-backs $\iota^\star\left[\mathcal{L}_{rheonomic}\right]$
of the time-honored \textit{rheonomic lagrangian} encode all
possible formulations of the same supersymmetric theory that are
either cohomologous or correspond to different cohomology classes.
\par
In view of this in the next section we provide an ultra short review
of integral forms.
\section{Integral Forms}
Integral forms are the crucial ingredients in order to define a
geometric integration theory on supermanifolds that inherits all the
good properties of differential form integration theory in
conventional geometry.
\par
We consider a supermanifold with $n$ bosonic dimensions and $m$
fermionic dimensions denoted by ${\cal M}^{(n|m)}$. We denote the
local coordinates in an open set as $\left ( x^a , \theta^\alpha
\right)$.
\par
A $(p|q)$ integral form $\omega^{(p|q)}$ has the following structure
\begin{eqnarray}
\label{genA} \omega^{(p|q)} = \omega(x,\theta) dx^{a_1}\dots
dx^{a_r} d\theta^{\alpha_1} \dots d\theta^{\alpha_s} \delta^{(b_1)}
(d\theta^{\beta_{1}}) \dots  \delta^{(b_{q})}(d\theta^{\beta_{q}})
\end{eqnarray}
where the $d\theta^\a$ appearing in the product are independent of
those appearing in the delta's $\delta(d\theta^\beta)$. By
$\omega(x,\theta)$ we denote the set of superfields provided by the
collection of the components $\omega_{[a_1 \dots a_r](\alpha_1 \dots
\alpha_s)[\beta_1 \dots \beta_q]}(x,\theta)$.
\par
The two quantum numbers $p$ and $q$ correspond to the {\it form}
number and the {\it picture} number and they range from $-\infty$ to
$+\infty$ for $p$ and $0 \leq q \leq m$.  The index on the delta
$\delta^{(a)}(d\theta^\a)$ denotes the degree of the derivative of
the delta function. The total picture of $\omega^{(p|q)}$
corresponds to the number of delta functions (we call it a {\it
superform} if $q=0$, an {\it integral form} if $q=m$, otherwise it
is dubbed a {\it pseudoform}). The total form degree is given by $p
= r + s - \sum_{i=1}^{i=q} b_i$ since the derivatives act
effectively as negative forms and the delta functions do not carry
any form degree. We recall the following properties
\begin{eqnarray}
\label{genB} d \delta^{(a)}(d\theta^\a) = 0\,, ~~~~~~ d\theta^\alpha
\delta^{(a)}(d\theta^\a) = - a \delta^{(a-1)}(d\theta^\a)\,, ~~ a
>0\,, ~~~~~ d\theta^\alpha \delta(d\theta^\a) = 0
\end{eqnarray}
The index $\alpha$ is not summed.
The indices $a_1 \dots a_r$ and $\beta_1 \dots \beta_q$ are anti-symmetrized, the
indices $\alpha_1 \dots \alpha_s$ are symmetrized because of the rules
\begin{eqnarray}
\label{genC} &&dx^a dx^b = - dx^b dx^a\,, ~~~ dx^a d\theta^\alpha =
d\theta^\alpha dx^a\,, ~~~
d\theta^\alpha d\theta^\beta= d\theta^\beta d\theta^\a\,, \nonumber \\
&& \delta(d\theta^\a) \delta(d\theta^\beta)  = -
\delta(d\theta^\beta)  \delta(d\theta^\alpha) \,, ~~~ dx^a
\delta(d\theta^\a) = \delta(d\theta^\alpha) dx^a\,, ~~~ d\theta^\a
\delta(d\theta^\beta) = \delta(d\theta^\beta) d\theta^\a\,.
\nonumber
\end{eqnarray}
\par
One can calculate the integral of an integral form $\omega^{(n|m)}$, it it is a top form, namely
an element of the line bundle known as the Berezinian bundle (the transition functions are represented
by the superdeterminat of the Jacobian)
and it can be locally expressed as
\begin{eqnarray}
\label{genE}
\omega^{(n|m)}  =  \omega(x,\theta) dx^1 \dots dx^n \delta(d\theta^1) \dots \delta(d\theta^m)\,.
\end{eqnarray}
where $\omega(x,\theta)$ is a superfield. By replacing the 1-forms
$dx^a, d\theta^\a$ as it follows $dx^a \rightarrow E^a = E^a_m dx^m
+ E^a_\mu d\theta^\mu$ and $d\theta^\alpha \rightarrow E^\alpha =
E^\alpha_m dx^m + E^\alpha_\mu d\theta^\mu$, we get
\begin{eqnarray}
\label{genEA}
\omega \rightarrow {\rm sdet}( E)  \, \omega(x,\theta) dx^1 \dots dx^n \delta(d\theta^1) \dots \delta(d\theta^m)
\end{eqnarray}
where $ {\rm sdet}(E)$ is the superdeterminant of the supervielbein $(E^a, E^\a)$.
\par
As in  conventional geometry, the integral form $\omega^{(n|m)}$ is
also viewed as a section of the cotangent bundle $T^* {\cal
M}^{(n|m)}$ and we perform the integral as follows
\begin{eqnarray}
\label{genD} I[\omega] = \int_{{\cal M}^{(n|m)}} \omega^{(n|m)} =
\int_{T^* {\cal M}^{(n|m)}} \omega(x, \theta, dx, d\theta) [dx
d\theta d(dx) d(d\theta)]
\end{eqnarray}
where the order of the integration variables is kept fixed. The symbol
$[dx d\theta d(dx) d(d\theta)]$ denotes the integration variables and it is invariant under
any coordinate transformations. The integrations over $d\theta$ and $d(dx)$ are Berezin integrals,
and $dx$ and $d(d\theta)$ are usual Lebesgue integrals. See Witten \cite{Witten:2012bg} for a complete discussion
on the symbol $[dx d\theta d(dx) d(d\theta)]$.

Let us consider a superform $\omega^{(n|0)}$ with degree $n$ equal
to the bosonic dimension of the bosonic submanifold ${\cal M}^{(n)}
\subset {\cal M}^{(n|m)}$. The superform is obtained in the usual
manner using the supervielbeins $E$, superconnections $\Omega$ and
the covariant derivatives of differential superforms, and formally
it can be integrated over the submanifold ${\cal M}^{(n)}$. However,
the transformation properties become manifest if the integral can be
converted into an integral form integrated over the entire
supermanifold. That can be achieved by constructing the Poincar\'e
dual form ${\mathbb Y}^{(0|m)}$ of the immersion of $\iota: {\cal
M}^{(n)} \longrightarrow {\cal M}^{(n|m)}$. Then, we can write the
integral as follows
\begin{eqnarray}
\label{genE}
I[\omega] = \int_{{\cal M}^{(n)}}  \omega^{(n|0)}_\star =
\int_{{\cal M}^{(n|m)}} \omega^{(n|0)} \wedge {\mathbb Y}^{(0|m)}
\end{eqnarray}
where $ \omega^{(n|0)}_\star  = \iota^* \omega^{(n|0)}$ is the pull-back of $ \omega^{(n|0)}$
on ${\cal M}^{(n)}$. Any variation of the embedding of  ${\cal
M}^{(n)}$ into ${\cal M}^{(n|m)}$ is compensated by a
diffeomorphism. The r.h.s. is the integral of an integral form for
which the usual rules of Cartan calculus do apply. Notice that we
can modify the embedding by changing ${\mathbb Y}^{(0|m)}$ by exact
terms if $\omega^{(m|0)}$ is a closed form. The Poincar\'e dual
${\mathbb Y}^{(0|m)}$ is closed and not exact. Any variation of the
embedding is exact: $\delta {\mathbb Y}^{(0|m)} = d \eta^{(-1|m)}$.
\par
For rigid supersymmetric models, the closed form $\omega^{(n|0)}$ is
represented by the Lagrangian of the model ${\cal L}^{(n|0)}(\Phi,V,
\psi)$ built using the rheonomic rules and it contains the dynamical
fields $\Phi$ (each dynamical field is promoted to a superfield) and
the rigid supervielbeins $V^a = dx^a + \theta \gamma^a d\theta\,,
\psi^\alpha = d\theta^\a$ satisfying the Maurer-Cartan equations
\begin{eqnarray}
\label{genEA} d V^a = \frac{i}{2} \psi \, \wedge \,\gamma^a \psi \,,
~~~~~d \psi^\a=0\,.
\end{eqnarray}
In the present formula, we have used real Majorana spinors; the notation can be made more precise only upon the choice
of the  dimensions $(n|m)$ of the supermanifold ${\cal SM}$.
\par
On the other hand the Poincar\'e dual form ${\mathbb Y}^{(0|m)}$
(a.k.a. {\it Picture Changing Operator} PCO in string theory
literature) contains only geometric data (for instance the
supervielbein or just the coordinates). One can choose a different
PCO which has some manifest symmetries.
\par
For rigid supersymmetric models we have
\begin{eqnarray}
\label{genF}
S_{rig} = \int_{{\cal M}^{(n|m)}} {\cal L}^{(n|0)}(\Phi, V, \psi) \wedge {\mathbb Y}^{(0|m)}(V, \psi)
\end{eqnarray}
with $d  {\cal L}^{(n|0)}(\Phi, V, \psi)  =0$ in order to change the PCO by exact terms.
The action ${\cal L}^{(n|0)}(\Phi, V, \psi)$ depends upon the dynamical fields $\Phi$ of the theory and
the PCO depends upon the geometrical data of the supermanifold.
\par
In the case of supergravity, the supervielbein $V^a$ and $\psi^\a$
are promoted to dynamical fields $(E^a, E^\alpha)$ so that the
action becomes:
\begin{eqnarray}
\label{genG}
S_{sugra} = \int_{{\cal M}^{(n|m)}} {\cal L}^{(n|0)}(\Phi, E) \wedge  {\mathbb Y}^{(0|m)}(E).
\end{eqnarray}
The closure of the action and the closure of the PCO implies the
conventional constraints of supergravity that reduce the independent
fields to the physical ones.

\section{Basics of D=10 $\mathcal{N}=1$ SYM}
We list some of the ingredients for D=10, $\mathcal{N}=1$
super-Yang-Mills theory. The theory is a maximally supersymmetric
model in D=10 and its field content consists of a gauge field
$\mathcal{A}_a(x)$ and of its superpartner $\chi_\a(x)$. In ten
dimensions, they are respectively assigned to the vector and to the
spinor representation of the Lorentz group. This amounts to $10 +
16$ off-shell degrees of freedom. The gauge field is defined up to a
gauge transformation which removes one degree of freedom, but still
7 bosonic dof.s are missing for an off-shell matching. On the other
hand, imposing the equations of motion, the gauge field reduces to 8
effective dof.s (since one additional constraint is removed by the
Hamiltonian constraint) while the gaugino reduces to 8 dof's since
the Dirac equation halves the off-shell ones. Hence there is a
physical supersymmetric on-shell multiplet\cite{Nahm:1977tg}.
\par
We assume an unspecified, compact, non-abelian gauge group ${\cal
G}$ and both the gauge field and the gaugino are assigned to the
adjoint representation of ${\cal G}$. The consistent equations of
motion are \cite{Brink:1976bc,Gates:1986tj}
\begin{eqnarray}
\label{SysA} \nabla^a \mathcal{F}_{ab} = [\chi, \gamma_b \chi]\,,
~~~~~ \gamma^a \nabla_a \chi = 0
\end{eqnarray}
where the bracket $[\cdot, \cdot]$ is with respect to the Lie
algebra of ${\cal G}$ while $\nabla_a$ denotes the covariant
derivative with respect to the gauge field in the adjoint
representation. These equations of motion are gauge and
supersymmetric covariant. They can be derived from a component
action which can be written as follows:
\begin{eqnarray}
\label{SysB} S = \int d^{10}x \, \, {\rm Tr}\left( - \frac14
\mathcal{F}^{ab} \mathcal{F}_{ab} + \chi \gamma^a \nabla_a
\chi\right)
\end{eqnarray}
the trace being taken in the adjoint representation of the gauge
group. The supersymmetry transformations \cite{Brink:1976bc} are the
following ones:
\begin{eqnarray}
\label{SysC} \delta \mathcal{A}_a =  \epsilon \gamma_a \chi\,, ~~~~~
\delta \chi_\alpha = \frac14 (\gamma^{ab} \epsilon)_\alpha \,\,
\mathcal{F}_{ab}
\end{eqnarray}
Their anticommutator closes on the translations, on the gauge
transformations and on the fermionic equations of motion
(\ref{SysA}). The supersymmetry algebra does not close off-shell
since there is no consistent auxiliary field set in the present
formulation.
\par
We remark that in the case of D=10, $\mathcal{N}=1$ SYM, there is
not a finite set of auxiliary fields that can be used to construct
an off-shell formulation and therefore a superspace description.
There have been several attempts (see \cite{Berkovits:1997wj} for a
clear discussion on this point) to build a superspace action for
this theory. Nonetheless, in \cite{DAuria:1981zjr} the existence has
been shown of a rheonomic action, whose associated equations of
motion in superspace lead to the supersymmetry transformations rules
and to the field equations of the component fields. That action, as
it will be explained in forthcoming sections, contains additional
off-shell degrees of freedom, precisely the 0-form superfields
$\mathfrak{F}_{ab}(x,\theta),  W^\a(x,
\theta)$. The first component of the former 
$\mathfrak{F}_{ab}(x,\theta=0)$ is identified, by its own field
equation, with the field strength $\mathcal{F}_{ab}(x)$ of the gauge
field $\mathcal{A}_a(x)$ while the first component of the latter is similarly 
identified with the gaugino field strength. 
Indeed the rheonomic formulation is intrinsically a first order
action -- no Hodge dual being used -- and therefore it needs these
additional superfields. As explained in \cite{DAuria:1981zjr} the
presence of these first order superfields circumvents the usual no-go
theorems of Siegel and Ro\v{c}ek \cite{Siegel:1981dx}.
\par
As it is well-known, reducing the superspace from $\mathcal{N}=1$,
$D=10$ down to $\mathcal{N}=4$, $D=4$ one finds the maximally
extended supersymmetric Yang--Mills theory. Equivalently, also in
the case of that theory no off-shell formulation is known that
makes the full $\mathcal{N}=4$ supersymmetry manifest. A superspace
action in the usual sense is still lacking  and maybe it does not
exist at all. Once more, in \cite{DAuria:1981zjr} the dimensional
reduction of the rheonomic action was obtained, yielding a long
expression which can be utilized as starting point in our new
integral form constructions.
\par
\section{Superspace  D=10 $\mathcal{N}=1$ SYM}
Even though there is no off-shell closure of the supersymmetry
algebra and the superspace action is absent, still we can  give a
superspace formulation of the equations of motion.
\par
We start from a super $1$-form $A^{(1|0)} = A_a V^a  + A_\alpha
\psi^\a$, (where the superfields $A_a(x,\theta)$ and
$A_\alpha(x,\theta)$ take value in the adjoint representation of the
gauge group) and we define the field strength
\begin{eqnarray}
F^{(2|0)} &\equiv& d A^{(1|0)} +  A^{(1|0)}
\wedge A^{(1|0)}\nonumber\\
& = & F_{ab} V^a \wedge V^b + F_{a\alpha} V^a \wedge \psi^\alpha +
F_{\alpha\beta} \psi^\a\wedge \psi^\beta\,, \label{SM-A}
\end{eqnarray}
where we have introduced the following field strengths (we recall
that $A_a$ is a bosonic superfield while $A_\a$ is a fermionic
superfield, therefore the notation $[A_a, A_b]$ denotes the Lie
commutator, while $\{A_\alpha, A_\beta\}$ denotes the anticommutator
to take into account the statistic of the superfields)
\begin{eqnarray}
\label{SM-B}
F_{ab} &=& \partial_a A_b - \partial_b A_a + [A_a, A_b]\,, ~~ \nonumber \\
F_{a\alpha} &=& \partial_a A_\alpha - D_\alpha A_a + [A_\alpha, A_b]\,, ~~ \nonumber \\
F_{\alpha\beta} &=& D_{(\alpha} A_{\beta)} + \gamma^a_{\alpha\beta}
A_a + \{A_\alpha, A_\beta\}\,.
\end{eqnarray}
In order to reduce the redundancy of degrees of freedom contained in the two
components $A_a$ and $A_\a$ of the $(1|0)$ connection, one imposes (by hand) the conventional constraint
\begin{eqnarray}
\label{SM-C} \iota_\alpha \iota_\beta F^{(2|0)} =0\,   ~~~
\Longleftrightarrow ~~~ F_{\alpha\beta} = \nabla_{(\alpha}
A_{\beta)} + \gamma^a_{\alpha\beta} A_a =0\,,
\end{eqnarray}
from which it follows that
\begin{eqnarray}
\label{SM-BA} A_a = - \frac{1}{8} \gamma_a^{\alpha\beta}
\nabla_{\alpha} A_\b\,, ~~~
W^\alpha = \nabla^\beta\nabla^\alpha A_\b\,, ~~~ 
F_{a\alpha} = \gamma_{a, \alpha\beta} W^\beta\,, ~~~~ F_{ab} =
(\gamma_{ab})_\a^{~\beta} \nabla_\beta W^\a\,,
\end{eqnarray}
and the dynamical equation
\begin{eqnarray}
\label{SM-BAA} \gamma_{[abcde]}^{\alpha\beta} \nabla_{(\alpha}
A_{\beta)} =0\,,
\end{eqnarray}
(where $\gamma_{[abcde]}^{\alpha\beta}$ is the anti-symmetrized
product of five gamma matrices) which implies the field equations
$\mathcal{N}=1, D=10$ super-Yang-Mills theory
\begin{eqnarray}
\label{SM-BB} \nabla^a F_{ab} = 0\,, ~~~~~~
\gamma^a_{\alpha\beta} \nabla_a W^\beta =0\,.
\end{eqnarray}

The gaugino field strength $W^\a$ is gauge invariant under the
non-abelian transformations $\delta A_\alpha = \nabla_\alpha
\Lambda$. These transformations follow from $\delta A = \nabla \Lambda$ where $\Lambda$ is a
$(0|0)$-form. The field strengths satisfy the following Bianchi's
identities
\begin{eqnarray}\label{SM-V}
&&\nabla_{[a} F_{bc]} =0 \,, ~~~~~ 
\nabla_\alpha F_{ab} + (\gamma_{[a} \nabla_{b]} W)_\alpha =0\,, \nonumber\\
&&F_{ab} + \frac12 (\gamma_{ab})^\alpha_{~\beta} \nabla_\alpha W^\beta =0 \,, ~~~~~~~~~
\nabla_\alpha W^\alpha =0\,.
\end{eqnarray}
\newcommand{\sint}{\int\!\!\!\!\!\!-}
and, by expanding the superfields $A_a, A_\a$ and $W^\a$ at the first level,  we
have
\begin{eqnarray}
\label{SM-VA} A_\alpha = (\gamma^a \theta)_\alpha \mathcal{A}_a(x) +
\chi_\a \frac{\theta^2}{2}\,, ~~~~~~~ A_a = \mathcal{A}_a(x) + \chi
\gamma_a \theta + \dots \,, ~~~~~~ W^\alpha = \chi^\alpha +
{F}^\alpha_{~\beta} \theta^\beta+ \dots\,,
\end{eqnarray}
where $\mathcal{A}_a(x)$ is the gauge field, $\chi_\a(x)$ is the
gaugino and $F_{\alpha\beta} = \gamma_{\alpha\beta}^{ab}
\mathcal{F}_{ab}$ is the gauge field strength with $\mathcal{F}_{ab}
=
\partial_a \mathcal{A}_b - \partial_b \mathcal{A}_a$.
\par
The Bianchi identities, together with the constraint
$F_{\alpha\beta} =0$ fix all components of the superfield $A_\a$ in
terms of the gauge field $\mathcal{A}_a(x)$ and of the gaugino
$\chi^\alpha(x)$ with the requirement that they are on-shell.
Therefore, the higher components of the superfield $A_\a$ are
completely fixed. This is equivalent to say that there is no
superspace off-shell formulation.

\section{Pure Spinor Formulation}
In the pure spinor formulation
\cite{Berkovits:2000fe,Berkovits:2001rb,Berkovits:2000nn} for superstrings and superparticles,
one starts from two--dimensional worldsheet fields or from
one--dimensional worldline fields $(x^a, \theta^\a, p_\a)$, representing the coordinates of the target space,
toghether with a pair of commuting spinor ghost fields $(\lambda^\a, w_\a)$. Then, one defines a BRST
differential operator $Q$ which restricted to zero modes reads
\begin{eqnarray}
\label{psA} Q = \lambda^\alpha D_\a
\end{eqnarray}
with $D_\a = \frac{\partial}{\partial \theta^\a} - (\gamma^a \theta)_\a \partial_a$
the superderivative. \footnote{To be
precise, $Q= \oint dz \lambda^\a(z) d_\a(z)$, where $\lambda^\a(z)$
is a commuting holomorphic worldsheet field and $d_\a(z)$ is the
holomorphic generator of fermionic constraints of Green-Schwarz
formulation of superstring. Acting on zero modes, $d_\a(z) \sim
D_\a$. } The latter satisfies the
commutation relations $\{D_\alpha, D_\b\} = -
2\gamma^a_{\alpha\beta}\partial_a$ and therefore $Q$ is nilpotent if
and only if \cite{Berkovits:2000fe,Berkovits:2001rb,Howe:1991mf,Howe:1991bx}
\begin{eqnarray}
\label{psAA} \lambda^\a \gamma^a_{\a\b} \lambda^\b =0\,, ~~~~~ a=0, \dots, 9\,.
\end{eqnarray}
These are the celebrated {\it pure spinor constraints} which can be
solved in terms of 11 independent complex degrees of freedom. The pure spinor
constraints (\ref{psAA}) are first class constraints and generate the gauge symmetry $\delta w_\a
= \eta_a (\gamma^a \lambda)_\a$ on the conjugated fields $w_\a$ and $\eta^\a$ is the gauge parameter 
associated to the first class constraints (\ref{psAA}). 
\par
In the superstring/superparticle formulation, the target space gauge
field and its superpartner are states of the appropriate Hilbert
space and they are described by a vertex operator $U^{(1)}$ at ghost
number one. Since only the pure spinor $\lambda^\a$ carries positive
ghost charge, a Lorentz invariant vertex operator has the generic
form
\begin{eqnarray}
\label{psBAA} U^{(1)} = \lambda^\alpha A_\a(x,\theta)
\end{eqnarray}
where $A_a(x,\theta)$ is a superfield. No vectorial partner
of $A_\a$ has been introduced and $A_\a$ is identified with the spinoral part of the
superconnection $A^{(1|0)}$. Acting with $Q$ on $U^{(1)}$, using
the algebras of superderivatives and imposing  BRST closure we get
\begin{eqnarray}
\label{psCA} \{Q, U^{(1)} \} + \{U^{(1)}, U^{(1)}\} = \frac12
\lambda^\alpha \lambda^\beta(D_\alpha A_\beta+ D_\beta A_\alpha +
\{A_\alpha, A_\b\}) =0\,.
\end{eqnarray}
In the above equation we added the second term in order to implement
the non-abelian gauge symmetry where the product $\{\cdot,
\cdot\}$ is the Lie algebra anticommutator between two
anticommuting vertex operators $U^{(1)}$. In the following, this
term is justified  as derived from an action of Chern-Simons type.
\par
Projecting the product $\lambda^\alpha \lambda^\b$ and using the
pure spinor constraints along the 5-form $\lambda \gamma^{[a_1 \dots
a_5]} \lambda$  yields the superspace equations of motion
\begin{eqnarray}
\label{psDAA} \gamma_{[a_1 \dots a_5]}^{\alpha\beta}(D_\alpha
A_\beta+ D_\b A_\alpha + \{A_\alpha, A_\b\})= 0\,.
\end{eqnarray}
If the vector part of $A^{(1|0)}$ is defined as
\begin{eqnarray}
\label{psDAB}
A_a =- \frac18 \gamma_a^{\a\b} D_\a A_\b\,,
\end{eqnarray}
the Bianchi identities imply all the identities derived in the
previous section and the correct equations of motion.
\par
Since $Q$ is a nilpotent
differential, the vertex operator is a physical quantity if
not BRST exact, or equivalently, if it is defined up to gauge
transformations
\begin{eqnarray}
\label{psE}
\delta U^{(1)} = Q \Lambda^{(0)} + [U^{(1)}, \Lambda^{(0)}] \,,
\end{eqnarray}
where $\Lambda^{(0)}$ is a superfield with vanishing ghost number.
Imposing a convenient gauge fixing for
the connection $A_\a$ ({\it e.g.} the Wess-Zumino gauge),
with an iterative procedure, one reconstructs all superfields
from the on-shell target space fields $\mathcal{A}_a(x), \chi_\a(x)$ (see \cite{Harnad:1985bc} for a complete discussion).
\par
The equations of motion (\ref{psDAA}) suggest a variational principle of the
form
\begin{eqnarray}
\label{psFAA}
S = \int d^{10}x d^{16}\theta d^{11}\lambda \mu(\theta, \lambda)
{\rm Tr}\left( \frac12 U^{(1)} Q U^{(1)} + \frac13 U^{(1)} U^{(1)} U^{(1)} \right)
\end{eqnarray}
where the trace is taken over the gauge group representation of the connection $A_\a$ which appears in the vertex
operator $U^{(1)}$ (\ref{psBAA}) and the multiplication of the three vertex operators in (\ref{psFAA}) is
matrix multiplication of the vertices at the same point in superspace. The integration
over the pure spinor is done with a suitable measure $\mu(\theta, \lambda)$ given by
\begin{eqnarray}
\label{psG} \mu(\theta, \lambda) &=& \theta^{\alpha_1} \dots
\theta^{\alpha_{11}} \epsilon_{\alpha_1 \dots \alpha_{16}}
(\gamma^{abc})^{\alpha_{12} \alpha_{13}} \gamma_a^{\alpha_{14}
\beta_1} \gamma_b^{\alpha_{15} \beta_2}
\gamma_c^{\alpha_{16} \beta_3} \iota_{\beta_1} \iota_{\beta_2} \iota_{\beta_3} \delta^{11}(\lambda) \nonumber \\
&=& (\epsilon \theta^{11})_{\alpha_1 \dots \alpha_5} T^{[\alpha_1
\dots \alpha_5] (\beta_1 \dots \beta_3)} \iota_{\beta_1}
\iota_{\beta_2} \iota_{\beta_3} \delta^{11}(\lambda)
\end{eqnarray}
where $\iota_\beta = \partial/ \partial \lambda^\b$. The tensor
$T^{[\alpha_1 \dots \alpha_5] (\beta_1 \dots \beta_3)}$ is the only
invariant tensor constructed from Dirac matrices with five
antisymmetrized spinorial indices and three symmetrized spinorial
indices.  This measure involves both the pure spinors $\lambda^\a$
and the fermionic coordinates $\theta^\a$. Since in $\mu(\theta, \lambda)$ 
there are already 11 $\theta$'s, the Berezin integration in (\ref{psFAA}) picks up 5
additional $\theta$'s from the expression in the bracket. The Dirac delta
functions $\delta^{11}(\lambda)$ are needed to integrate over the
pure spinors: they are bosonic variables and the integral has to
be convergent. The measure $\mu(\theta, \lambda)$ is BRST invariant since
\begin{eqnarray}
\label{psGA} Q \mu(\theta, \lambda) &=& 11\,  \lambda^{\alpha_1}
\theta^{\alpha_2} \dots \theta^{\alpha_{11}}
\epsilon_{\alpha_1 \dots \alpha_{16}} T^{[\alpha_{11} \dots \alpha_{16}] (\beta_1 \dots \beta_3)}  \iota_{\beta_1} \iota_{\beta_2} \iota_{\beta_3} \delta^{11}(\lambda) \nonumber \\
&=& 11\,  \theta^{\alpha_2} \dots \theta^{\alpha_{11}}
\epsilon_{\alpha_1 \dots \alpha_{16}} T^{[\alpha_{11} \dots
\alpha_{16}] (\beta_1 \dots \beta_3)}
\delta^{\alpha_1}_{(\beta_1}  \iota_{\beta_2} \iota_{\beta_3)} \delta^{11}(\lambda) \nonumber \\
&=&  33 \,  \theta^{\alpha_2} \dots \theta^{\alpha_{11}}
\epsilon_{\alpha_1 \dots \alpha_{16}} T^{[\alpha_{11} \dots
\alpha_{16}] (\alpha_1 \beta_2 \beta_3)} \iota_{\beta_2}
\iota_{\beta_3}\delta^{11}(\lambda) =0\,.
\end{eqnarray}
The last equality followos from the properties of $T^{[\alpha_1
\dots \alpha_5] (\beta_1 \dots \beta_3)}$, where a further
antisymmetrization of one of its spinorial indices $\beta_1,
\beta_2$ or $\beta_3$ makes it vanishing. The second line is due to
the fact that the $\lambda^{\alpha_1}$ produced by the BRST
variation of $\theta^{\alpha_1}$ has to be annihilated by one of the
derivatives $\iota_\b$ acting on $\delta^{11}(\lambda)$ otherwise it
vanishes (integration-by-parts).
\par
Integrating over the pure spinor space we get
\begin{eqnarray}
 S = \int d^{10}x d^{16}\theta (\epsilon
\theta^{11})_{\alpha_1 \dots \alpha_5} (\gamma^{abc})^{\alpha_{1}
\alpha_{2}} \gamma_a^{\alpha_{3} \beta_1} \gamma_b^{\alpha_{4}
\beta_2} \gamma_c^{\alpha_{5} \beta_3} {\rm Tr}\left( \frac12
A_{\beta_1} D_{\beta_2} A_{\beta_3} +  \frac13 A_{\beta_1}
A_{\beta_2} A_{\beta_3} \right)\nonumber\\
\label{psH}
\end{eqnarray}
where we have used integration by parts for the pure spinors,
removing them from the vertex operators, and the Dirac delta's
$\delta^{11}(\lambda)$ to integrate over them. The Berezin integral
over the $\theta$'s requires  extraction of  the five $\theta$'s
terms from the bracket. Collecting all the pieces, one gets the
action (\ref{SysB}), namely the component action. The expression
(\ref{psH}) can be regarded as a superspace action.

\subsection{Pure Spinor Volume form and PCO's}

As shown in \cite{Berkovits:2000fe}, in the pure spinor formalism it
exists a ghost-number 3 cohomology class admitting the following
representative
\begin{eqnarray}
\label{psB}
\omega^{(3|0)} = (\lambda \gamma^a \theta)
(\lambda \gamma^b \theta) (\lambda \gamma^c \theta) (\theta \gamma_{abc} \theta)
\end{eqnarray}
which is BRST invariant modulo pure spinor constraints  and it is not exact.
The expression is not supersymmetric invariant since it explicitly depends upon the $\theta$'s, nonetheless
its variation is BRST exact: $\delta_\epsilon  \omega^{(3|0)} = d \Sigma^{(2|0)}$ where
$\delta_\epsilon \theta^\a = \epsilon^\a$ and $\delta_\epsilon \lambda^\a=0$.
 \par
What about the volume form? As pointed out in
\cite{Berkovits:2004px}, by introducing the delta functions of the
pure spinors $\delta(\lambda)$,
 the volume form is an integral form (restricted to $\theta$'s and $\lambda$'s space) reads
\begin{eqnarray}
\label{psD} {\rm Vol}^{(0|11)} = \epsilon_{\alpha_1 \dots
\alpha_{16}} \theta^{\alpha_1} \dots  \theta^{\alpha_{16}}
\delta^{11}(\lambda)\,.
\end{eqnarray}
where we defined the expression
\begin{eqnarray}
\label{psDA}
\delta^{11}(\lambda) = \epsilon_{\alpha_1 \dots
\alpha_{11}} \delta(\lambda^{\alpha_1}) d \lambda^{\alpha_1}\wedge
\dots \wedge \delta(\lambda^{\alpha_{11}}) d \lambda^{\alpha_{11}}
\end{eqnarray}
taking into account only the independent degrees of freedom.
Another way to write the same quantity is by introducing a set of
commuting spinors $C_{\alpha, i}$ with $i=1, \dots, 11$ and writing
\begin{eqnarray}
\label{psDA} \bigwedge_{i=1}^{11} \delta(C_{\alpha, i} \lambda^\a) d
(C_{\alpha, i} \lambda^\a) =  \delta^{11}(\lambda)
\end{eqnarray}
since the determinant coming from the differentials cancels against that coming from the Dirac delta functions.
This is normalized as
\begin{eqnarray}
\label{psDB}
\int_{{\cal M}^{(0|16)}} {\rm Vol}^{(0|11)} =1\,.
\end{eqnarray}
The volume form (\ref{psD}) is BRST closed and it is not exact. It has ghost number
zero and maximal picture.
\par
According to the theory of integral forms \cite{Castellani:2014goa,Castellani:2015paa},
 we can construct the PCO, Hodge dual to $\omega^{(3|0)}$ ($\star: H^{(3|0)} \rightarrow H^{(-3|11)}$)
 \cite{Catenacci:2016qzd}.
It should belong to the space $H^{(-3|11)}$, such that locally it
satisfies
\begin{eqnarray}
\label{psE}
\mathbb{Y}^{(-3|11)} \wedge \omega^{(3|0)}  = {\rm Vol}^{(0|11)} \,,
\end{eqnarray}
The closure and non-exactness of $\omega^{(3|0)}$ and of $ {\rm Vol}^{(0|11)}$ imply
 that $ Q \mathbb{Y}^{(-3|11)}  =0$.

The result is the following
\begin{eqnarray}
\label{psF} \mathbb{Y}^{(-3|11)} = \epsilon_{\alpha_1 \dots
\alpha_{16}} \theta^{\alpha_1} \dots  \theta^{\alpha_{11}}
\gamma_a^{\alpha_{12} \beta_1}  \gamma_b^{\alpha_{13} \beta_2}
\gamma_c^{\alpha_{14} \beta_3} (\gamma^{abc})^{\alpha_{15}
\alpha_{16}} \frac{\partial}{\partial \lambda^{\beta_1}}
\frac{\partial}{\partial \lambda^{\beta_2}} \frac{\partial}{\partial
\lambda^{\beta_3}} \delta^{11}(\lambda)\,. 
\end{eqnarray}
By using the pure spinor properties and gamma matrix algebra, this
PCO is indeed $Q$-closed and not $Q$-exact. Eq. (\ref{psE}) can be
easily checked: one can note that the fermionic coordinates match
the total number and the derivatives on the delta's act by
integration-by-parts on $\omega^{(3|0)}$ absorbing the three
$\lambda$'s.


\section{The rheonomic Lagrangian of D'Auria--Fr\'e--Da Silva}
As announced in the introduction, one can write the action of $D=10$
super Yang--Mills theory in the geometrical language of rheonomy.
This was done in \cite{DAuria:1981zjr}. The independent
fields are $\mathfrak{F}_{ab}(x, \theta), W^\a(x,\theta)$ and the
connection $A^{(1|0)}= A_a(x, \theta) V^a + A_\a(x,\theta)$, while
the supervielbein $E^A = (V^a, \psi^\a)$ is kept constant (when
coupling this action to supergravity, the vielbein $E^A$ becomes
dynamical). The starting point to build the rheonomic action is
provided by the component action (\ref{SysB}), by the weights of the
different fields, and by Lorentz invariance. The action ${\cal
L}^{(10|0)}$ is a $(10|0)$ superform and it is built avoiding the
Hogde dual product. As constructed in \cite{DAuria:1981zjr} and
reviewed in \cite{castdauriafre} it reads
\begin{eqnarray}
\label{SYMA} {\cal L}^{(10|0)} &=& \left( - \frac{1}{90}
\mathfrak{F}_{ab} \mathfrak{F}^{ab} \, V^{a_1}{}_\wedge \dots
{}_\wedge V^{a_{10}} +
\mathfrak{F}^{a_1 a_2}\, {F}^{(2|0)}{}_\wedge V^{a_3} \dots{}_\wedge V^{a_{10}} \right. \nonumber \\
&&\left. + 2 i \mathfrak{F}^{a_1 a_2} W \gamma_{a} \psi{}_\wedge
V^a{}_\wedge V^{a_3} \dots {}_\wedge V^{a_{10}} + \frac{4}{9} i W
\gamma^{a_1} \nabla W {}_\wedge V^{a_2} \dots {}_\wedge V^{a_{10}}
 \right.\nonumber \\
&&\left. + \frac{8}{3} i W \gamma^{a_1 \dots a_3} \psi {}_\wedge
{F}^{(2|0)}{}_\wedge  V^{a_4} \dots {}_\wedge
V^{a_{10}}\right.\nonumber\\
&&\left. +\left(1+\frac 38 a\right)\, W \gamma^{a_1 \dots a_3} W
\psi {}_\wedge \gamma_a \psi {}_\wedge V^a {}_\wedge V^{a_4}_\wedge
\dots {}_\wedge V^{a_{10}}
\right.\nonumber \\
&&\left. + a \, W \,\gamma^{a_1 a_2 b} \, W \, \psi {}_\wedge
\gamma_b \psi {}_\wedge V^{a_3}_\wedge \dots {}_\wedge
V^{a_{10}}\right)\,
\epsilon_{a_1 \dots a_{10}}\nonumber\\
&& - 84 i \left( A^{(1|0)} {}_\wedge {F}^{(2|0)} - \frac13 A^{(1|0)}
{}_\wedge  A^{(1|0)} {}_\wedge  A^{(1|0)}\right){}_\wedge \psi
{}_\wedge\gamma^{a_1 \dots a_5} \psi {}_\wedge V_{a_1}{}_\wedge
\dots {}_\wedge  V_{a_5}
\end{eqnarray}
with ${F}^{(2|0)} = d A^{(1|0)} + A^{(1|0)} {}_\wedge A^{(1|0)}$,
satisfying  the Bianchi identity $d {F}^{(2|0)} + A^{(1|0)}
{}_\wedge {F}^{(2|0)} =0$. The variation of the action with respect
to  the $(0|0)$-forms $\mathfrak{F}_{ab}, \, W^\a$  yields the
following constraints:
\begin{eqnarray}
\label{SYMB} {F}^{(2|0)} = \mathfrak{F}_{ab} V^a {}_\wedge V^b - 2 i
W \gamma_a \psi {}_\wedge V^a\,, ~~~~~~~ \nabla W = V^a \nabla_a W -
\frac14 \gamma^{ab} \psi \,\,\mathfrak{F}_{ab} \,,
\end{eqnarray}
which imply the equations of motion
\begin{eqnarray}
\label{SYMC} \nabla^a \mathfrak{F}_{ab} = 0\,, ~~~~~~ \gamma^a
\nabla_a W =0\,,
\end{eqnarray}
Comparing eq.s (\ref{SYMB}) with eq.s (\ref{SM-A}), we see that the 
field equation of the superfield $\mathfrak{F}_{ab}$ is algebraic
and implies:
\begin{equation}\label{implicone}
   F_{ab}\, = \,  \mathfrak{F}_{ab}
\end{equation}
and
\begin{equation}\label{impliconebis}
    {F}_{a\beta} \, = \, - \,2 i
(W \gamma_a)_\beta\,.
\end{equation}
Hence on the mass--shell the superfield $\mathfrak{F}_{ab} =
\mathcal{F}_{ab}(x) + {\cal O}(\theta) $ starts with the bosonic
field strength while $W^\alpha = \chi^\a(x) + {\cal O}(\theta)$
starts with the gaugino.
\par
The action depends upon a free parameter $a$ which parameterizes the
two different spinorial structures. It is easy to see that by
setting $\psi=0$, the action reduces to the first order formalism
for the component action of SYM as given in (\ref{SysB}).
\par
The truly interesting term is the last one. It has the form {\it
Chern-Simons $\times$ a Chevalley-Eilenberg cohomology class}, which
is common to all rheonomic formulations of supersymmetric theories
and it plays a crucial role here. Indeed, using  Fierz identities it
can be proved  that
\begin{eqnarray}
\label{SYMD} \omega^{(7|0)} = \psi {}_\wedge \gamma^{a_1 \dots a_5}
\psi {}_\wedge   V_{a_1}{}_\wedge   \dots {}_\wedge  V_{a_5} \,,
\end{eqnarray}
is an element of Chevalley-Eilenberg cohomology for the
$\mathcal{N}=1$ super--Poincar\'e Lie algebra in $D=10$. The
\textit{superform} $\omega^{(7|0)}$ is closed but not exact. From
the point of view of supergravity and of Free Differential
Algebras\footnote{For a modern review of the theory of FDA.s in
relation with Sullivan's theorems and Chevalley-Eilenberg cohomology
of super Lie Algebras see sections 6.3 and 6.4 (pages 227-238) in
the second volume of \cite{maiobuk}.}, this Chevalley-Eilenberg
cohomology class is responsible for the extension of the
$\mathcal{N}=1$, $D=10$ super Poincar\'e Lie Algebra to an FDA
including  a 6-form ${B}^{[6]}$ which can be regarded as the
magnetic dual of the Kalb-Ramond $2$-form ${B}^{[2]}$ (see
\cite{DAuria:1987qjh} about this point).
\par
By varying the action with respect to the independent fields
$\mathfrak{F}_{ab}, W^\a, A^{(1|0)}$, we get the following equations
of motion in superspace:
\begin{eqnarray}
\label{SYMDA1} &-&\frac{1}{45}\epsilon_{a_1
 \dots  a_{10}}\, \mathfrak{F}_{ab} \,  \, V^{a_1}
\,\wedge \dots \wedge\, V^{a_{10}}
 + \, \epsilon_{a b a_3 \dots  a_{10}}
F^{(2|0)} \, \wedge \,
 V^{a_3} \,\wedge \dots \wedge\, V^{a_{10}}  \nonumber \\
&+& 2i \, \epsilon_{a b a_3 \dots a_{10}} \, W \gamma_c \psi \,
\wedge \, V^c \, \wedge \, V^{a_3} \,\wedge \dots \wedge\,
V^{a_{10}} =0\, ,
\end{eqnarray}
\begin{eqnarray}
\label{SYMDA2} &+& \frac{8 i}{9}\epsilon_{a_1 a_2\dots  a_{10}}\,
(\gamma^{a_1} \nabla W)\, \wedge \,
  V^{a_2} \,\wedge \dots \wedge\,
V^{a_{10}}\nonumber\\
 &+& 2 i\, \epsilon_{a_1 a_2 b a_3 \dots a_{10}} \, \mathfrak{F}^{a_1 a_2} (\gamma^b
 \psi)\, \wedge \,
 V^{a_3} \,\wedge \dots \wedge\, V^{a_{10}} \nonumber \\
&-& 2 \, \epsilon_{a_1 a_2
 \dots  a_{10}}\, (\gamma^{a_1} W) (\psi \, \wedge \, \gamma^{a_2} \psi) \, \wedge \,  V^{a_3} \,\wedge
\dots \wedge\,
V^{a_{10}}\nonumber\\
& + &\frac{8 i }{3} \epsilon_{a_1 a_2 a_3 a_4 \dots  a_{10}}
(\gamma^{a_1 a_2 a_3} \psi)\, \wedge \, F^{(2|0)} \, \wedge \,
 V^{a_4} \,\wedge \dots \wedge\, V^{a_{10}} \nonumber \\
&+& 2 \left(1 + \frac{8}{3} a\right)\epsilon_{a_1 a_2 a_3 a_4 \dots
a_{10}} (\gamma^{a_1 a_2 a_3} W) (\psi \, \wedge \,  \gamma_{b}
\psi)\, \wedge \,  V^b \, \wedge \, V^{a_4} \,\wedge \dots \wedge\,
V^{a_{10}}
\nonumber \\
&+& 2 a \epsilon_{a_1 a_2 a_3  \dots a_{10}} \, (\gamma^{a_1 a_2 b}
W) (\psi \, \wedge \, \gamma_{b} \psi)\, \wedge \,  V^{a_3} \,\wedge
\dots \wedge\, V^{a_{10}} =0\,,
\end{eqnarray}
\begin{eqnarray}
\label{SYMDA3} & + &\epsilon_{a_1 a_2 a_3  \dots  a_{10}} \nabla
\mathfrak{F}^{a_1 a_2}\, \wedge \,    V^{a_3} \,\wedge \dots
\wedge\, V^{a_{10}} \nonumber\\
&+& 4 i \epsilon_{a_1 a_2 a_3 a_4 \dots  a_{10}} \mathfrak{F}^{a_1
a_2} (\psi \, \wedge \,\gamma^{a_3} \psi)\, \wedge \,
 V^{a_4} \,\wedge \dots \wedge\, V^{a_{10}} \nonumber \\
&+& \frac{8 i}{3} \epsilon_{a_1  \dots a_3 a_4 \dots a_{10}}(\nabla
W \, \wedge \,\gamma^{a_1  \dots a_3} \psi)\, \wedge \, V^{a_4}
\,\wedge \dots \wedge\,
V^{a_{10}}\nonumber\\
& + &\frac{8 i}{3} \epsilon_{a_1  \dots a_3 a_4 a_5 \dots a_{10}} (W
\gamma^{a_1  \dots  a_3} \psi) \, \wedge \, (\psi \, \wedge \,
\gamma^{a_4} \psi)\, \wedge \,
 V^{a_5} \,\wedge \dots \wedge\, V^{a_{10}} \nonumber \\
&-& 84 i\, F^{(2|0)}\, \wedge \,  \psi \, \wedge \, \gamma^{a_1
\dots a_5} \psi \,\wedge \, V_{a_1}\, \wedge   \, \dots \, \wedge \,
V_{a_5} =0\,.
\end{eqnarray}
Notice that the first two equations are $(10|0)$ superforms and the
last is a $(9|0)$ superform. By expanding $F^{(2|0)}$ into $(2|0)$
components and $\nabla W$ into $(1|0)$ components one finds the
already anticipated constraints (\ref{SYMB}) which determine the
supersymmetry transformation rules. Notice that these constraints
reproduce only the on-shell supersymmetry transformations and from
those one recovers also the dynamical equations of motion
\begin{eqnarray}
\label{SYMC} \nabla^a F_{ab} = 0\,, ~~~~~~ \gamma^a \nabla_a W =0\,,
\end{eqnarray}
The above field equations implied from the superspace constraints
follow also from the projection of the field equations on the
maximal vielbein sector ($\psi = 0$ projection). This is the
fundamental consistency check that guarantees the supersymmetry of
the component lagrangian as extracted from the rheonomic one. These
equations are indeed necessary in order to satisfy the set of
equations (\ref{SYMDA1})-(\ref{SYMDA3}). Notice that expanding the
$(10|0)$ forms into the different components, at the highest order
in the gravitino  field $\psi$ ( the $\psi^3 V^7$ sector, in this
case), one sees that only the algebraic properties of gamma matrices
are needed to solve such equations. Note that there are several
redundancies among the equations. This is due to the fact that at
each order in the gravitino expansion, the action encodes the same
amount of information. This allows us to extract different
superspace actions from the rheonomic action (\ref{SYMA}).

\section{PCO's and interpolating actions}
Here we start from the rhonomic action (\ref{SYMA}) and we show how
to interpolate between the spacetime action (\ref{SysB}) and the
superspace action of the form (\ref{psFAA}).
\par
In order to derive the component action, we have to integrate it over the supermanifold ${\cal M}^{(10|16)}$
and for that we need a PCO which multiplies ${\cal L}^{(10|0)}$ of (\ref{SYMA}). This is achieved by
constructing the expression
\begin{eqnarray}
\label{SYME}
\mathbb{Y}^{(0|16)} = \theta^{16} \delta^{16}(\psi)
\end{eqnarray}
where we recall that $\psi = d\theta$ for flat superspace. This
projects to the space with $\theta^\a =0$ and $\psi^\a=0$.
\par
Then, we obtain
\begin{eqnarray}
\label{SYMF}
S &=& \int_{{\cal M}^{(10|16)}} {\cal L}^{(10|0)} \wedge \mathbb{Y}^{(0|16)} =
\int
\left( - \frac{1}{90} \mathfrak{F}_{ab} \mathfrak{F}^{ab} \, V^{a_1}{}_\wedge \dots {}_\wedge V^{a_{10}} \right. \nonumber \\
&+& \left. \mathfrak{F}^{a_1 a_2}\, {F}^{(2|0)}{}_\wedge V^{a_3}
\dots{}_\wedge V^{a_{10}} + \frac{4}{9} i W \gamma^{a_1} \nabla W
{}_\wedge V^{a_2} \dots {}_\wedge V^{a_{10}}
\right) \epsilon_{a_1 \dots a_{10}} \wedge \mathbb{Y}^{(0|16)} \nonumber \\
&=& \int d^{10}x \left( - \frac{1}{4} \mathcal{F}_{ab}
\mathcal{F}^{ab} + \chi \gamma^{a} \nabla_a \chi \right)
\end{eqnarray}
where $\mathcal{F}_{ab}$ depends only on $x$, as already stated and
$\chi^\a = W^\a(x, \theta =0)$. Note that all the vielbeins $V^a$ reduce
to $dx^a$ since $\theta = d\theta =0$. Furthermore, in the last step
we have substituted back into the lagrangian the algebraic field
equation of the auxiliary $0$-form $\mathfrak{F}_{ab}\mid_{\theta=0}
\, = \, \mathcal{F}_{ab}$. In this way we obtain the standard second
order form of the component lagrangian.
\par
Let us now change the above PCO and let us introduce a new choice
\begin{eqnarray}
\label{SYMG} \mathbb{Y}^{(0|16)}_{p.s.} = V^{a_1} {}_\wedge V^{a_2}
{}_\wedge V^{a_3} {}_\wedge V^{a_4} {}_\wedge V^{a_5}
\epsilon_{\b_1 \dots \b_{16}} \theta^{\b_1} \dots
\theta^{\b_{11}} (\gamma_{a_1} \iota)^{\b_{12}} \dots
(\gamma_{a_5} \iota)^{\b_{16}} \delta^{16}(\psi)
\end{eqnarray}
Notice that the above PCO has zero form degree (since it contains  5
vielbeins and 5 contractions $\iota_\a$) and picture equal to
sixteen. Notice that it contains only eleven $\theta$'s. They carry
an upper index $\a$ while the contraction $\iota_\a$ has a lower
index. Therefore, the combination $(\gamma_a \iota)^\a$ cannot be
contracted with $\theta^\a$. Then, one needs the Levi-Civita tensor
$ \epsilon_{\b_1 \dots \b_{16}}$ which leaves eleven
anti-symmetric indices to be contracted with the $\theta$'s.  Notice
that $\iota_\a$ is a commuting differential operator, the
combination  $(\gamma_a \iota)^\b$ works as an anticommuting
quantity because of the factor $V^{a_1} {}_\wedge V^{a_2} {}_\wedge
V^{a_3} {}_\wedge V^{a_4} {}_\wedge V^{a_5}$ which is antisymmetric in
the vector indices $a_1\dots a_5$.
\par
Let us check the closure of $\mathbb{Y}^{(0|16)}_{p.s.}$. By using $\psi = d\theta$,
we have
\begin{eqnarray}
\label{checkA} d \mathbb{Y}^{(0|16)}_{p.s.} &=& 5 ( \psi
\gamma^{a_1} \psi) V^{a_2} {}_\wedge V^{a_3} {}_\wedge V^{a_4}
{}_\wedge V^{a_5} \epsilon_{\b_1 \dots \b_{16}}
\theta^{\b_1} \dots \theta^{\b_{11}}
(\gamma_{a_1} \iota)^{\b_{12}} \dots (\gamma_{a_5} \iota)^{\b_{16}} \delta^{16}(\psi) \nonumber \\
&+& 11\, V^{a_1} {}_\wedge V^{a_2} {}_\wedge V^{a_3} {}_\wedge
V^{a_4} {}_\wedge V^{a_5}  \epsilon_{\b_1 \dots \b_{16}}
\psi^{\b_1} \dots \theta^{\b_{11}}
(\gamma_{a_1} \iota)^{\b_{12}} \dots (\gamma_{a_5} \iota)^{\b_{16}} \delta^{16}(\psi) \nonumber \\
&=&  10   V^{[a_2} {}_\wedge V^{a_3} {}_\wedge V^{a_4} {}_\wedge
V^{a_5} (\gamma_{[a_1} \gamma^{a_1]} \gamma_{a_2})^{\b_{12}
\b_{13}} (\theta^{11} \epsilon)_{\b_{12} \dots \b_{16}}
(\gamma_{a_3} \iota)^{\b_{14}} \dots (\gamma_{a_5]} \iota)^{\b_{16}} \delta^{16}(\psi) \nonumber \\
&+& 55 V^{a_1} {}_\wedge V^{a_2} {}_\wedge V^{a_3} {}_\wedge V^{a_4}
{}_\wedge V^{a_5} \gamma_{a_1}^{\b_{11} \b_{12}}
(\theta^{10} \epsilon)_{\b_{11} \dots \b_{16}} (\gamma_{a_2}
\iota)^{\b_{13}} \dots (\gamma_{a_5} \iota)^{\b_{16}}
\delta^{16}(\psi)\,,\nonumber \\
&=& 0\,.
\end{eqnarray}
where $ (\theta^{11} \epsilon)_{\b_{12} \dots \b_{16}} = 
\epsilon_{\b_1 \dots \b_{11}\b_{12} \dots \b_{16}} \theta^{\b_{1}} \dots \theta^{\b_{11}}$. 

In eq. (\ref{checkA}), the first line vanishes since the matrix $(\gamma_{a_1}
\gamma^{a_1} \gamma_{a_2})^{\b_{12} \b_{13}}$ has to be
antisymmetric in the spinorial indices and this implies that it
should be proportional to the gamma matrix $\gamma_{a_1 a_2 a_3}$
which is totally antisymmetric in the vectorial indices. However,
the contractions between Lorentz indices in the formula imply that
this vanishes. The second line vanishes since the indices of the
symmetric gamma matrix $\gamma_{a_1}$ are contracted with the
$\epsilon$-tensor in spinorial space. Therefore this implies that
the PCO is indeed closed.
\par
Before completing the discussion about the action, we observe the following
relation
\begin{eqnarray}
\label{checkB}
\omega^{(7|16)} &=& \omega^{(7|0)}\wedge \mathbb{Y}^{(0|16)}_{p.s.} \nonumber \\
&=& V^{10} (\theta^{11} \epsilon)_{\alpha_1 \dots \alpha_5}
T^{[\alpha_1 \dots \alpha_5](\beta_1 \beta_2 \beta_3)}
\iota_{\beta_1} \iota_{\beta_2} \iota_{\beta_3} \delta^{16}(\psi)
\end{eqnarray}
where $T^{[\alpha_1 \dots \alpha_5](\beta_1 \beta_2 \beta_3)} =
(\gamma^{abc})^{[\alpha_1 \a_2}  (\gamma_a)^{\alpha_3(\beta_1}
\gamma_b^{\alpha_4 \beta_3} \gamma_c^{\a_5] \b_3)}$ which is the
invariant spinorial numerical tensor discussed in (\ref{SysB}) and
$V^{10} = \epsilon_{a_0 \dots a_9} V^{a_0} \wedge \dots \wedge
V^{a_9}$ is the bosonic volume form. Notice that the new integral
form $\omega^{(7|16)}$ has all desired  good properties. It is
Lorentz invariant and it is closed since both $\omega^{(7|0)}$ and
$\mathbb{Y}^{(0|16)}_{p.s.}$ are closed.
\par
\par
Now, we apply the PCO to the action. Due to the presence of five
vielbeins $V^a$, all terms except the last one drop out and we are
left with
\begin{eqnarray}
\label{SYML}
S = \int_{{\cal M}^{(10|16)}} \left[\left( {\cal A} {}_\wedge {\cal F}
- \frac13  {\cal A} {}_\wedge  {\cal A} {}_\wedge  {\cal A}\right)\wedge \,
\psi \gamma^{a_1 \dots a_5} \psi {}_\wedge   V_{a_1}{}_\wedge   \dots {}_\wedge  V_{a_5}\right] \wedge
 \mathbb{Y}^{(0|16)}_{p.s.}
\end{eqnarray}
Notice that the five contractions $\iota_\a$ appearing in
$\mathbb{Y}^{(0|16)}$ absorb five $\psi$'s both from
$\omega^{(7|0)}$ and from the Chern-Simons term. In particular,
the two $\psi$'s from $\omega^{(7|0)}$  and three $\psi$'s
are removed from the Chern-Simons action. Since the latter is a three form, this means
that it selects ${\cal L}^{(3|0)} = \psi^\alpha \psi^\b \psi^\gamma
A_{(\alpha} D_\beta A_{\gamma)} + \dots$ (as in the pure spinor
action). The vielbeins $V^a$ are ten in total, hence  they arrange
themselves into a  scalar. The Dirac delta's allow for the
integration over the $\psi$'s so that  we are finally left with the
counting of $\theta$'s. Notice that since we have already eleven
$\theta$'s in the PCO, we need to take 5 $D$-derivatives of the
action. This finally yields the SYM action in components.
\par
In conclusion, the action admits the following very elegant writing:
\begin{eqnarray}
\label{SYMMA} S = \int_{{\cal M}^{(10|16)}}
\left[\underbrace{\mbox{Tr}\,\left( A^{(1|0)} {}_\wedge {F}^{(2|0)}
- \frac13 A^{(1|0)} {}_\wedge A^{(1|0)} {}_\wedge
A^{(1|0)}\right)}_{\mbox{gauge CS form}}\wedge \,\omega^{(7|0)}
\wedge
 \mathbb{Y}^{(0|16)}_{p.s.} \right]
\end{eqnarray}
\par
The combination $ \omega^{(7|0)} \wedge
 \mathbb{Y}^{(0|16)}_{p.s.}$ yields
 \begin{eqnarray}
\label{SYMN}
 \omega^{(7|0)} \wedge
 \mathbb{Y}^{(0|16)}_{p.s.} = V^1 \wedge \dots \wedge V^{10} \wedge \mathbb{Y}^{(-3|11)}
\end{eqnarray}
where $\mathbb{Y}^{(-3|11)}$ is given in (\ref{psF}).
\par
We succeeded to show that the pure spinor formulation and the rheonomic formulation
of SYM produce the same superspace action. The rheonomic formulation
has its origin into the theory of integral forms which shares common features with pure
spinor superstring by the introduction of the target-space PCO's. We hope that this new point of view
might be used for a fruitful re-formulation of type IIB supergravity along the same lines
\cite{castella-pesando,Sen:2015nph}.


\section{Conclusions}
We add some considerations to the above presented discussion and we
single out some interesting  problems for future investigations.
\begin{description}
\item[a)] We have illustrated the similarities between the pure spinor formulation and
the geometrical formulation based on the rheonomic action plus the
crucial ingredient of the PCO's (\textit{i.e.} Poincar\'e duals of
\textit{top bosonic cycles} in superspace). It would be desirable to
establish a dictionary between the two frameworks. We are tempted to
identify the pure spinor $\lambda^\a$ (in D=10), satisfying the
constraints $\lambda^\a \gamma^a_{\alpha\beta} \lambda^\beta=0$ with
the differential $\psi^\a = d\theta^\a$ although the latter are not
constrained.
\par
Thus, a ``trial" dictionary might be
\begin{eqnarray}
\label{psA} \theta^\alpha \leftrightarrow \theta^\a\,, ~~~~~~~
\lambda^\alpha  \leftrightarrow d\theta^\a\,, ~~~~~~~ Q
\leftrightarrow  d\,,
\end{eqnarray}
The form number is replaced by the ghost number.
\par
Since the pure spinor components are treated as bosonic commuting
coordinates, their Dirac delta functions correspond to  usual Dirac
delta's. Then, in order to identify $\delta(\psi)$ with the pure
spinor Dirac delta, we set
\begin{eqnarray}
\label{psC}
\delta(\psi^\a) = \delta(\lambda^\a) d\lambda^\a
\end{eqnarray}
where the index $\alpha$ is not summed. Notice that in this way, the
l.h.s. has no form degree, while the $\delta(\lambda^\a)$ has
negative form degree $(-1)$ to compensate the form degree of
$d\lambda^\a$. In addition, since the usual Dirac delta function
$\delta(\lambda^\a)$ is a commuting quantity, the resulting
$\delta(\psi^\a)$ is anticommuting, as it should be for our
purposes. The combination $\delta(\lambda^\a) d\lambda^\a$ is
trivially BRST closed.
\item[b)] The second remark is to point out that the rheonomic action is not $d$ closed (without the use of
the superspace constraints that reduce the theory on shell).
According to  common lore \footnote{see \cite{castdauriafre} and
\cite{maiobuk} for reviews.} and to  actual calculations, this is
intimately related to the absence of an auxiliary field formulation
for an off-shell extension of the algebra. The present derivation
does not solve such an issue, yet we emphasize that starting from a
completely different point of view we arrive at  the same result
(\ref{SYMMA}) in both formalisms.
\item[c)]In relation with point b) let us also stress that in
\cite{DAuria:1981zjr} Bianchi identities of $D=10$ super Yang-Mills
theory in the anholonomic setup, proper to the rheonomic approach,
were analyzed up to the third order in the spinor derivatives in
search for a possible set of auxiliary fields. No match was found
between bosons and fermions up to that order. The conjecture
therefore was put forward that the only off-shell representation is
that containing \textit{819200 bosons + 819200 fermions} which
corresponds to no constraint whatsoever on the \textit{superform}
$A^{(1|0)}$. Indeed starting from the most general parametrization
of the curvature superform in terms of coefficient superfields
arranged into irreducible $\mathrm{SO(1,9)}$ representations
\footnote{The affix ${}^{(n)}$ appended to each coefficient
superfield denotes the dimension of the corresponding Lorentz
representation.}:
\begin{eqnarray}
  F^{(2|0)} &=& F_{ab} \, V^a \wedge V^b - 2 \psi \, \left(\xi_a^{(144)} - i \, \gamma_a \zeta^{(16)} \right)
  \wedge V^a\nonumber \\
   & & +i \, \mathcal{B}^{(10)}_a \, \psi \, \wedge \, \gamma^a \psi \, + \, i
   \mathcal{B}^{(126)}_{a_1,\dots a_5} \, \psi \, \wedge \, \gamma^{a_1\,
   \dots a_5} \, \psi \label{piripicchio}
\end{eqnarray}
it was shown that any Lorentz covariant constraint, namely the
suppression of any of the four representations
$\mathbf{144}$,$\mathbf{16}$,$\mathbf{10}$ or $\mathbf{126}$
immediately reduces the space-time fields to the mass shell. This
left the possibility that the Lorentz covariant constraint could be
imposed by suppressing some of the Lorentz irreducible
representation in the spinor derivatives of the above or in the
derivatives of the derivatives. Yet, up to third derivatives, as we
recalled, no viable constraint was found and this motivated the
quoted conjecture. Now, in view of the identification in eq.
(\ref{psA}) a new line of thinking is brought to attention. That no
Lorentz covariant off-shell representation does exist, except the
largest one, is probably a correct conclusion in unconstrained
superspace,  yet what about \textit{constrained superspace} in line
with the ideas utilized in \cite{Fre:2007xy}? Let us observe that if
the theta's satisfy the Lorentz covariant constraint $d\theta \wedge
\gamma^a\, d\theta \, = \, 0$, analogous to the pure spinor
constraint, then we also have $\psi \wedge\,  \gamma^a \psi =0$ and
in eq. (\ref{piripicchio}) the superfield $\mathcal{B}^{(10)}_a$
naturally disappears. What happens in the subsequent orders is
matter to be anlyzed, but it is not inconceivable that a consistent
set of auxiliary fields can now be found. This is a matter for
future analysis.
\item[d)] The two PCO's (\ref{SYME})  and (\ref{SYMG}) are two possible choices;  there
are several other that are cohomologous to them and it remains to be
explored which kind of action they might  lead to.
\end{description}

\section*{Acknowledgement} 
We thank C. Maccaferri, L. Castellani and R. Catenacci for fruitful discussions.

\newpage


\begin{thebibliography}{10}

\bibitem{Nahm:1977tg}
W.~Nahm, ``{Supersymmetries and their Representations},'' {\em Nucl.
Phys.},
  vol.~B135, p.~149, 1978.

\bibitem{Green:2012oqa}
M.~B. Green, J.~H. Schwarz, and E.~Witten, {\em {Superstring Theory
Vol. 1}}.
\newblock Cambridge University Press, 2012.

\bibitem{Green:2012pqa}
M.~B. Green, J.~H. Schwarz, and E.~Witten, {\em {Superstring Theory
Vol. 2}}.
\newblock Cambridge University Press, 2012.

\bibitem{Brink:1976bc}
L.~Brink, J.~H. Schwarz, and J.~Scherk, ``{Supersymmetric Yang-Mills
  Theories},'' {\em Nucl. Phys.}, vol.~B121, pp.~77--92, 1977.

\bibitem{Witten:1985nt}
E.~Witten, ``{Twistor - Like Transform in Ten-Dimensions},'' {\em
Nucl. Phys.},
  vol.~B266, pp.~245--264, 1986.

\bibitem{Gates:1986tj}
S.~J. Gates, Jr. and H.~Nishino, ``{On D = 10, N=1 Supersymmetry,
Superspace
  Geometry and Superstring Effects. 2.},'' {\em Nucl. Phys.}, vol.~B291,
  p.~205, 1987.

\bibitem{Berkovits:1997wj}
N.~Berkovits and C.~M. Hull, ``{Manifestly covariant actions for D =
4 selfdual
  Yang-Mills and D = 10 superYang-Mills},'' {\em JHEP}, vol.~02, p.~012, 1998.

\bibitem{Harnad:1985bc}
J.~P. Harnad and S.~Shnider, ``{Constraints and field equations for
  ten-dimensional super Yang-Mills theory},'' {\em Commun. Math. Phys.},
  vol.~106, p.~183, 1986.

\bibitem{Howe:1991mf}
P.~S. Howe, ``{Pure spinors lines in superspace and ten-dimensional
  supersymmetric theories},'' {\em Phys. Lett.}, vol.~B258, pp.~141--144, 1991.
\newblock [Addendum: Phys. Lett.B259,511(1991)].

\bibitem{Howe:1991bx}
P.~S. Howe, ``{Pure spinors, function superspaces and supergravity
theories in
  ten-dimensions and eleven-dimensions},'' {\em Phys. Lett.}, vol.~B273,
  pp.~90--94, 1991.

\bibitem{Berkovits:2000fe}
N.~Berkovits, ``{Super Poincare covariant quantization of the
superstring},''
  {\em JHEP}, vol.~04, p.~018, 2000.

\bibitem{Berkovits:2001rb}
N.~Berkovits, ``{Covariant quantization of the superparticle using
pure
  spinors},'' {\em JHEP}, vol.~09, p.~016, 2001.


\bibitem{Berkovits:2004px}
  N.~Berkovits,
  {``Multiloop amplitudes and vanishing theorems using the pure spinor formalism for the superstring,''}
  JHEP {\bf 0409}, 047 (2004)
  doi:10.1088/1126-6708/2004/09/047
  [hep-th/0406055].

\bibitem{castdauriafre}
L.~Castellani, R.~D'Auria, and P.~Fr\'e, {\em Supergravity and
superstrings: A
  Geometric perspective. Vol. 1,2,3.}
\newblock 1991.

\bibitem{DAuria:1981zjr}
R.~D'Auria, P.~Fr{\'e}, and A.~J. da~Silva, ``{Geometric Structure
of $N=1,\,
  D=10$ and $N=4, \, D=4$ Superyang-mills Theory},'' {\em Nucl. Phys.}, vol.~B196,
  pp.~205--239, 1982.

\bibitem{Fre:1981ny}
P.~Fr{\'e}, ``{N=1 supersymmetric Yang-Mills theory on the
supergroup manifold and the supercurrent},'' {\em Lett. Nuovo Cim.},
vol.~30, p.~507, 1981.

\bibitem{Fre:1981my}
P.~Fr{\'e}, ``{Remarks on the Supergroup Manifold Approach to
Supersymmetric
  {Yang-Mills} Theories and the Explicit Construction of the $N=2$ Case},''
  {\em Nucl. Phys.}, vol.~B187, pp.~376--388, 1981.

\bibitem{Castellani:2014goa}
L.~Castellani, R.~Catenacci, and P.~A. Grassi, ``{Supergravity
Actions with
  Integral Forms},'' {\em Nucl. Phys.}, vol.~B889, pp.~419--442, 2014.

\bibitem{Castellani:2015paa}
L.~Castellani, R.~Catenacci, and P.~A. Grassi, ``{The Geometry of
  Supermanifolds and New Supersymmetric Actions},'' {\em Nucl. Phys.},
  vol.~B899, pp.~112--148, 2015.

\bibitem{Castellani:2017ycm}
L.~Castellani, R.~Catenacci, and P.~A. Grassi, ``{Super Quantum
Mechanics in
  the Integral Form Formalism},'' 2017.

\bibitem{pappo1}
P.~Fr\'e and P.~A. Grassi, ``{The Integral Form of D=3 Chern-Simons
Theories
  Probing ${\mathbb C}^n/\Gamma$ Singularities}.'' arXiV/hep-th1705.00752,
  2017.

\bibitem{Catenacci:2016qzd}
  L.~Castellani, R.~Catenacci and P.~A.~Grassi,
  ``Integral representations on supermanifolds: super Hodge duals, PCOs and Liouville forms,''
  Lett.\ Math.\ Phys.\  {\bf 107}, no. 1, 167 (2017)
  doi:10.1007/s11005-016-0895-x
  [arXiv:1603.01092 [hep-th]].

\bibitem{Castellani:2016ibp}
L.~Castellani, R.~Catenacci, and P.~A. Grassi, ``{The Integral Form
of
  Supergravity},'' {\em JHEP}, vol.~10, p.~049, 2016.

\bibitem{Grassi:2016apf}
P.~A. Grassi and C.~Maccaferri, ``{Chern-Simons Theory on
Supermanifolds},''
  {\em JHEP}, vol.~09, p.~170, 2016.

\bibitem{Witten:2012bg}
E.~Witten, ``{Notes on Supermanifolds and Integration},'' 2012.

\bibitem{Siegel:1981dx}
W.~Siegel and M.~Rocek, ``{On off-shell supermultiplets},'' {\em
Phys. Lett.},
  vol.~105B, pp.~275--277, 1981.

\bibitem{Berkovits:2000nn}
N.~Berkovits, ``{Cohomology in the pure spinor formalism for the
  superstring},'' {\em JHEP}, vol.~09, p.~046, 2000.
\bibitem{maiobuk}
P.~G. Fr{\'e}, {\em Gravity, a Geometrical Course}, vol.~1,2.
\newblock Springer Science \& Business Media, 2012.
\bibitem{DAuria:1987qjh}
R.~D'Auria and P.~Fre, ``{Duality in Superspace and Anomaly Free
Supergravity:
  Some Remarks},'' {\em Mod. Phys. Lett.}, vol.~A3, p.~673, 1988.
\bibitem{castella-pesando}
  L.~Castellani and I.~Pesando,
  ``The Complete superspace action of chiral D = 10, N=2 supergravity,''
  Int.\ J.\ Mod.\ Phys.\ A {\bf 8}, 1125 (1993).
  doi:10.1142/S0217751X9300045X
\bibitem{Sen:2015nph}
  A.~Sen,
  ``Covariant Action for Type IIB Supergravity,''
  JHEP {\bf 1607}, 017 (2016)
  doi:10.1007/JHEP07(2016)017
  [arXiv:1511.08220 [hep-th]].
\bibitem{Fre:2007xy}
  P.~Fre and P.~A.~Grassi,
  ``Constrained Supermanifolds for AdS M-Theory Backgrounds,''
  JHEP {\bf 0801} (2008) 036
  doi:10.1088/1126-6708/2008/01/036
  [arXiv:0704.3413 [hep-th]].
\end{thebibliography}
\end{document}